\documentclass[11pt]{article}
\usepackage{amsmath}
\usepackage{latexsym}
\usepackage{amsfonts}
\usepackage[normalem]{ulem}
\usepackage{soul}
\usepackage{array}
\usepackage{amssymb}
\usepackage{extarrows}
\usepackage{graphicx}
\usepackage[svgnames,table]{xcolor}
\usepackage{enumitem}
\usepackage{multicol}
\usepackage{tikz}
\usepackage{changepage}
\usepackage{setspace}
\usepackage{ragged2e}
\usetikzlibrary{shapes.symbols,shapes.geometric,shadows,arrows.meta}
\usepackage[many]{tcolorbox}    	% for COLORED BOXES (tikz and xcolor included)

\tikzset{>={Latex[width=1.5mm,length=2mm]}}

\usepackage[utf8]{inputenc}
\newtcolorbox{boxB}{
    fontupper = \bf\color{black}, % font color
    boxrule = 1.5pt,
    colframe = red,
    rounded corners,
    arc = 5pt   % corners roundness
}

\begin{document}
\pagestyle{empty}

\begin{Center}
{\fontsize{20pt}{36pt}\selectfont \textcolor[HTML]{000000}{Expanding Horizons - Science White Paper}{\fontsize{30pt}{36.0pt}\selectfont \textcolor[HTML]{0070C0}{}\par}\par}

\vspace{3cm}

{\fontsize{15pt}{36pt}\selectfont \textcolor[HTML]{000000}{ Bridging stellar evolution and planet formation:\vspace{-0.6cm}\\from birth, to survivors of the fittest, to the second generation of planets}{\fontsize{30pt}{36.0pt}\selectfont \textcolor[HTML]{0070C0}{}\par}\par}
\end{Center}\par

% How planets meet stellar evolution. From survivors of the fittest to the second generation of planets.

% peering into the universality of dust physics and planet-host interaction
\vspace{1cm}

\textbf{Lead author}\\
Akke Corporaal, European Southern Observatory, Chile\\

\textbf{Co-authors/co-signatories}\\
Toon De Prins, KU Leuven, Belgium\\
L\'ea Planquart, Chalmers University of Technology, Sweden\\
Kateryna Andrych, Macquarie University, Australia\\
Narsireddy Anugu, CHARA Array of Georgia State University, United States \\
% Julien Drevon, European Southern Observatory, Chile \\
Devika Kamath, Macquarie University, Australia\\
Jens Kammerer, European Southern Observatory, Germany\\
Stefan Kraus, University of Exeter, United Kingdom\\
Foteini Lykou, Konkoly Observatory, Hungary\\ 
Alexis Matter, C\^ote d'Azur Observatory, France\\
Claudia Paladini, European Southern Observatory, Chile\\
Marie M. Rodr\'iguez S.,  Universidad de Chile, Chile\\ 
Hans Van Winckel, KU Leuven, Belgium\\

\newpage

\pagestyle{plain}

\textbf{Abstract:}\\
Stars and planets form, live, and evolve in unison. Throughout the life of a star, dusty circumstellar discs and stellar outflows influence the further evolution of both the star(s) and their orbiting planet(s).
Planet-forming discs, winds of red giant branch (RGB) or asymptotic giant branch (AGB) stars, and post-RGB/post-AGB discs are examples of such host environments where dust physics plays a key role. %were (proto)planets are expected.
The physical processes that occur during each of these stages establishes how the Solar System as well as exoplanetary systems were formed, are evolving, and will eventually die.
This White Paper aims to bridge the fields of stellar evolution and planet formation by peering into the dust kinematics and macrostructure formation, and its effect on planet-host interaction, in dusty environments from stellar birth to death.
Near-future advancements in the 2030s will enable the detection, orbital monitoring and atmospheric/mineralogical characterisation of close-in (proto)planets across diverse stages of stellar evolution. To take full advantage of these developments by the 2040s, we should develop the capabilities required to image the varied dusty environments in which planets are entrained over their lifetime. This will enable extensive testing of current theoretical understandings -- from the micro-scales of dust assembly to the deeply interlinked macro-scales of planet-host interactions -- across diverse settings often too small, distant, and faint to be resolved in the next decade, simultaneously providing valuable constraints on the two-way interplay of dusty host environments and planetary formation/evolution.

\vspace{-0.4cm}

\subsection{Introduction and current status}
\vspace{-0.7cm}
\begin{multicols}{2}

The formation and evolution of planets and their host stars is a key part of stellar and planetary astrophysics.
These fields eventually aim to decipher the physics that occurs during the birth, life, and death of the Solar System as well as exoplanetary systems of various architectures. 
From stellar birth to death, orbiting planets, their embryos, or their leftover materials interact with their host environment.
In the case of low-to intermediate-mass ($\leq$\,8\,M$_\odot$) stars, the most pronounced planet-host interactions are the dusty environments invoked during stellar formation and the (post-)red giant phases.
% They present us with challenging key questions in lo-mass ($<8\,stellar and planetary astrophysics.

\vspace{0.3cm}

(Proto)planets are thought to be the shaping mechanisms of various observed structures seen in circumstellar discs and stellar outflows, each in their characteristic way (Decin et al. 2020; Bae et al. 2023). 
While planet-forming discs arise as a natural by-product of star formation, unambiguously imaged protoplanets embedded in their natal birth environment are limited to two systems and three giant planets to date (PDS\,70bc: Keppler et al. 2018, Haffert et al. 2019; WISPIT\,2b: Close et al. 2025, van Capelleveen et al. 2025).
In the absence of direct signatures of planet formation processes in the planet-forming environment, indirect detection via observed dust substructures potentially caused by planet-disc interaction has been put forward instead (e.g.\ Andrews et al.\,2018; Maio et al. 2025).
Whether planet-induced or not, such substructures heavily affect dust grain processing. A key location in these respects is the water snowline, where condensation and diffusion of gas-phase water enhances the sticking properties of molecules, promoting planet formation by locally concentrating dust (Drazkowska \& Alibert 2017).
As the snowline is expected to move inward over the disc lifetime (${\leq}10$\,Myr), a snowline could create multiple ring-like structures in the dust distribution interior to the snowline (e.g.\ Owen 2020), depending on the stellar mass and the radial and vertical temperature profiles (Kennedy \& Kenyon 2008).
Moreover, current modelling efforts suggest that varying dynamical histories and chemical abundances in the birth environment may play an important role in the efficiency and primary locations of dust grain growth, which affect the process and outcome of planet formation (e.g.\ Nielsen et al. 2023).
However, our current understanding of planet formation is biased to the nearest star forming regions due to currently achievable sensitivities and angular resolutions.
\textbf{While our understanding of the physics and interaction of protoplanets with their natal discs is progressing fast using available multi-wavelength observing facilities, dedicated data processing, and advanced simulations, key questions regarding the timing, mechanisms, and locations of planetary formation and their dependency on environmental factors remain unanswered.}\vspace{0.3cm}

 % Besides metallicity, planet properties also correlate with stellar age, effective temperature and mass

% Current open questions in the formation of planetary systems are related to the the timing, mechanisms, and location of planetary formation.

As low-to intermediate-mass ($\leq$\,8\,M$_\odot$) stellar evolution progresses, the fate of the formed planetary systems remains illusive. 
In particular, planetary systems may be subjected to engulfment or orbital rearrangements following the expansion of the host star as it reaches the red giant branch (RGB) and asymptotic giant branch (AGB; Veras 2016).
Nevertheless, violent dynamics leading to planetary destruction during the later white dwarf phase remains the leading explanation of the detection of metal lines in the photospheres of white dwarfs, indicating that planets survive at least until this stage (e.g.\ Zuckerman et al. 2003, 2010; Veras et al. 2016; Coutu et al. 2019).
%as their host star(s) evolve(s) is limited.
To understand planetary survival around post-main sequence systems, stellar mass loss and mass transfer during and beyond the RGB and AGB are of particular interest.
% As planets are often found within 5\,astronomical units from the star
Mass loss during the AGB phase through dust-driven stellar winds or outflows has been suggested to be shaped by stellar or planetary companions (e.g.\ Maes et al. 2021). 
Moreover, substellar companions embedded in dusty clouds (former giant exoplanets now grown to a brown dwarf mass) are put forward as explanations of the long-period secondary period variables in AGB stars (Soszy\'nski et al. 2021).
During these giant stages of post-main sequence stellar evolution, mass transfer between binary stars often additionally leads to the formation of dusty circumstellar discs.
As a result of such disc formation and its subsequent evolution, surviving planets are expected to undergo orbital rearrangement through planet-disc interactions.
In particular, the presence of discs around post-RGB/post-AGB binaries has been observationally established (De Ruyter et al. 2006; Kluska et al. 2022).
These are even being considered as potential cradles of second-generation planet formation, given their Keplerian dynamics and complex inner and outer disc morphologies, including asymmetries and substructures (Sahai et al. 2011; Bujarrabal et al. 2013; Kluska et al. 2019; Gallardo Gava et al. 2021; Corporaal et al. 2023, Andrych et al. 2025). 
While no true planet detections have been made as of yet, these systems provide a window into circumstellar disc and dust physics in a different environmental parameter space, providing constraints highly complementary to the ones derived from planet-forming disc around young stars. 
Additionally, key open questions in evolved stellar physics regarding dust physics and its shaping of planetary systems remain. \textbf{In particular, it is unclear how we can link the RGB, AGB, and post-RGB/AGB phases in terms of dust, disc, and outflow physics, and how planetary systems are shaped and evolved in such dusty environments. This includes the mechanisms and timing of dust grain growth in the late stages of stellar evolution.}

 % if they manage to stay in a large enough orbit to avoid engulfment by the star, when the latter increases it's radius as it ascends the Red Giant Branch and Asymptotic Giant Branch (AGB). We have explored this hypothesis by determining: (1) the planet's survival during the AGB phase and the orbital changes due to the AGB mass-loss using the range of initial masses for white dwarfs progenitors, and (2) the range of parameters under which an outflow from the gas planet caused by irradiation by the planetary nebulae central star leads to the total destruction of the planet. We show that planets with masses less than one Jupiter mass do not survive the planetary nebula phase if located initially at orbital distances smaller than 3-5 AU. Planets more massive than two Jupiter masses around low-mass stars (1 M⊙ on the Main Sequence) survive the planetary nebulae stage down to orbital distances of ∼3 AU. Planets around white dwarfs with masses of MWD > 0.7 M⊙ are generally expected to be found at orbital radii r> 15 AU. 

\vspace{0.3cm}
The environments mentioned above cover various evolutionary stages -- from stellar formation, through the giant phases and all the way to the white dwarf stage -- and are found to be complex from small- (i.e.\,${<}1$\,au) to large- (100s of au) scales.
Such shaping processes start close to the star, at sub-astronomical unit scale and even within the first few stellar radii, where star(s) and possible close-in substellar companions(s) interact. The impact of these interactions then propagate to the large scales of the system.
% In addition to the diverse exoplanet population, detected asymmetries, substructures, and chemical complexities in circumstellar matter indicate that our theories and models should be able to account for this diversity. 
Connecting the results of studies from various stellar evolutionary stages, and covering a variety of environmental factors such as metallicities, is key to bridge stellar and planetary astrophysics.
These require targeted efforts dedicated to understanding the physical properties at very small spatial scales by both observational and theoretical means.
Ongoing work in this field, as well as expected progress in the next decade will set the stage for scientific exploration in dusty environments in the forthcoming decades.
This science white paper covers the following question: \\

\end{multicols}

\vspace{-0.6cm}

\begin{boxB}
\begin{center}
  \textit{How does dust processing from stellar birth to death regulate when, where, and how planets form and survive across the Hertzsprung–Russell diagram?}

\end{center}
\end{boxB}

% \vspace{-3.5cm}

 % complex and varied dynamical processes that influence planetary bodies after the star has turned off of the MS. 

% Efforts devoted to understanding the physical properties by both observational and theoretical means. A key facet to improve our knowledge of such a population is to connect existing findings obtained from the study of its various evolutionary phases, from formation, through the Main Sequence, and beyond the post-AGB phase.

% High angular resolution observations of stars and their circumstellar environment from the optical to the sub-millimetre have revealed the complexity of the circumstellar environments.

% The formation and long-term evolution of planets is yet to be understood.
% The impact of planets in the shaping of the circumstellar environment seems important although yet unclear.
% Current simulations show ...
% While Earth-like planets are expected are yet beyond the detectable ...
% The formation and evolution of Earth-like planets is one of the key questions. 

% Circumstellar discs are found across stars at different stellar evolutionary stages. 

% From both observations and simulations it becomes evident that there is no one-size-fits-all. Diversity exists in terms of the spatial distribution of gas and dust in the disc, the presence and nature of substructures, the strength of star-disc interaction, and their sizes.
% Binary systems/stellar multiples

% the physics of planet formation universal across the HR diagram and at different metallicities?}

% Infrared interferometric advancements
\vspace{-0.5cm}

\subsection{Expected 
advancement in the 2030s}
\vspace{-0.7cm}

\begin{multicols}{2}
In the next decade, several upcoming facilities, in synergy with current instrumentation, will make significant progress on this question.
On the one hand, the sensitivity of current instrumentation and infrastructures is consistently being pushed towards fainter limits. An example is the recent laser guide star facility at the Very Large Telescope Interferometer (VLTI; GRAVITY+ collaboration et al. 2025). 
Moreover, complementary approaches to resolve the close-by environment of exoplanets using high-contrast capabilities will be enabled by the upcoming VLTI/NOTT instrument (Defr\`ere et al. 2024).
% Moreover, complementary approaches to resolve and characterise the close-by environment of exoplanets and gaseous streams are being considered by using the upcoming high resolution spectroscopic ELT/ANDES instrument and the high-contrast capabilities of VLTI/NOTT instrument, respectively.
% ; Seidel et al. 2025)the high-contrast capabilities of the upcoming VLTI/NOTT instrument will allow for direct characterisation of the young exoplanet population around the snowline (Defr\'ere et al. 2024).
On the other hand, planned facilities will enable higher sensitivity (e.g.\ the ELT) and large-scale detections of planets (e.g.\ PLATO).
Opportunities include observations of the brighter sources in the SMC/LMC, allowing us to constrain the chemical inventory of circumstellar material in metal-poorer environments (using e.g. ELT/METIS).
Additionally, characterisation of close-by planets through high resolution spectroscopy will provide complementary constraints to the history and evolution of circumstellar material (using e.g. ELT/ANDES; Seidel et al. 2025).
Moreover, a significant increase of the detection of planets across the Hertzsprung-Russel diagram, including planets orbiting evolved stars, is  expected with the PLATO space mission (Rauer et al. 2014; 2025). 
% The wide-band and sensitivity upgrade (WSU

\vspace{0.3cm}
Despite these expected advancements, current and planned facilities will keep key windows into dust processing, including planet formation and evolution in dusty environments across the Hertzsprung-Russell diagram, inaccessible. 
In particular, the expected locations for key processes such as dust grain growth, dust clumping, and planetary interactions with the dusty environments in the aforementioned stellar evolutionary stages, remain at or beyond the spatial resolution limit provided by current and planned facilities.

% Lea: there is also a ongoing effort + future propsect with the ELT to characterise the zoo of exoplanet atmospheres and gaseous streams (eg. Seidel et al. Nature, 2025) through HR spectroscopy.

% This is a complementary approach to resolve close-by enviroment of exoplanets

% and hundreds of smaller Earth-like planets (R$_\mathrm{planet}\,{\leq}2\,\mathrm{R_{Earth}}$ from M dwarfs to G-type stars is 
% On the one hand the larger collecting area as provided by the ELT, especially the METIS instrument will allow to peer into the chemical inventory in the thermal infrared of the circumstellar environments from planet forming discs, to AGB stars, to post-AGB discs. The thermal infrared is of particular importance as these trace molecules such as water-ice, nano-diamonds and PAHs.
% The
% As such, this will allow to open up to different environments, needed to reduce the bias towards the brighter systems that we know of.

% Moreover, from a planet detection perspective, space facilities such as the PLATO mission (), are expected to detect planets across the Hertzsprung-Russel diagram, including planets orbiting evolved stars. This includes the smaller Earth-like planets that are currently sparse.

% The advancement of simulations ... 
\vspace{-0.5cm}
\subsection{Prospects for the 2040s}
\vspace{-0.3cm}

Mapping the kinematics and dust formation at currently unresolvable scales close to the host star is crucial for our understanding of dust physics, and thus for stellar and planetary formation and evolution.
In more practical terms, key degeneracies such as snowline- vs planet-driven substructures and companion- vs wind-shaped outflows cannot be broken without 0.1\,mas resolution at ${\sim}$1-3$\,\mu$m (e.g. water ice band), and 0.3\,mas at ${\sim}$10$\,\mu$m (e.g. silicate dust emission). The main opportunities that would be unlocked are the following:
\\
 % \uline{The smallest-scales of dust physics
 % across stellar evolution}
\vspace{-0.3cm}
\begin{itemize}[leftmargin=*]
\setlength\itemsep{-0.1cm}

    % \begin{itemize}
    \vspace{-0.5cm}
     \item Imaging the very inner regions (0.01\,au-10\,au) of planet-forming discs and post-RGB/post-AGB discs at 0.1\,mas scales would allow us to test and refine our understanding of macrostructure formation and planet-disc interaction.
     Key inner disc substructures crucial for planet formation that are not resolved today include the predicted effects of the ${\sim}$130\,K water snowline, dust sublimation (${\sim}$1000-1500\,K), and migration traps, peaking at wavelengths of ${\sim}\,1{-}3\,\mu$m (e.g. Lau et al. 2024).
     As these processes are expected to take place on 0.05-0.1\,au scales (depending on the stellar mass and the disc age (Kennedy \& Kenyon 2008; Flock et al. 2019)), these are not yet resolvable even for the nearest (${\leq}150$\,pc) planet-forming discs.
    % For young stars, dust sublimation, water snowlines, and planetary migration are expected at 0.1-10\,au (1-66\,mas) from the star, depending on the stellar mass and the disc age (Kennedy \& Kenyon 2008; Flock et al. 2019).
    % For post-RGB/post-AGB discs, these processes are expected at 1-10\,au (1-10\,mas) at their typical luminosities and distances of ${>}1$\,kpc (e.g. Kluska et al. 2019).
    % The expected 
   % The effects of the physics, such as gap opening as a result of planetesimal formation or dust traffic jams due to local condensation and diffusion (Drazkowska \& Alibert 2017) is not yet directly testable.
       Similarly, the imaging of substructures caused by rapid planetary migration in the disc would help us to understand orbital re-arrangements in circumstellar disc environments, including planet-forming discs and post-RGB/post-AGB discs.
     \item Resolving 0.1\,au scales in dusty outflows and circumstellar discs of RGB, AGB, and post-RGB/post-AGB stars (at typical distances of $>1$\,kpc) will allow us to perform mineralogical characterisation and probe planet-host interactions in evolved stars as well as potential second generation planet formation.
          % \vspace{-0.2cm}
    % \begin{itemize}
    %  \vspace{-0.35cm}
         \item Mapping dusty environments beyond the nearby star forming regions, AGB stars, and post-RGB/post-AGB discs, enabling the expansion of statistical studies by comparing structure, mineralogy, and evolution over a larger variety of environments (e.g.\ age, density, stellar mass, metallicity).
         The key here is to resolve dusty environments that are too faint or too distant to be resolved at sub-au scales today (e.g.\ young stars ($\leq$10\,Myr) at 1-1.5\,kpc, Prisinzano et al. 2022).
    \end{itemize}

% TODO: snowline explanation, how it creates substructures, test proposed physics, traffic jam,...
% % Calculation of hill sphere of a typical Jupiter, and explanation of why imaging is needed for it.
% Planetary substructures such as rings and gaps that are now commonly detected in the outer regions, likely as a result of dust clearing by forming giant planets, are also expected to be created for the terrestrial counterparts.
% As terrestrial planet formation is expected in the inner disc regions and its smaller planetary mass results in gaps on the order of ... (simulations needed, ref?)
% Plus explanation of what 0.1mas can do for dust coagulation, sublimation, ... 
% These require imaging capabilities at 0.1\,mas scales
% and the circumplanetary discs, which peak at mid-infrared wavelengths are are $\sim$ 0.1\,mas.

\textbf{This scientific outlook requires capabilities that are best provided by a future interferometric facility, operating from near-infrared to mid-infrared wavelengths ($\mathbf{\leq1}$ to $\mathbf{{>}10}$\,$\boldsymbol{\mathrm{\mu m}}$) with angular resolutions of ${\sim}$0.1\,mas} -- a factor of five better than reachable with the current sharpest eyes on the sky; the VLTI and CHARA arrays (currently providing imaging capabilities at angular scales up to 0.7-0.5\,mas, respectively).
Such angular resolution are only available at infrared wavelengths, with baselines ${\ge} 1$\,km.
The technological advancement would include an array that is scalable to a larger number of telescopes needed for truly robust snapshot imaging capabilities, as reliable interferometric image reconstructions require excellent spatial frequency coverage.
% \textbf{Technological advancements to reach precise interference at such baselines would include infrared heterodyne interferometry}, 
Such capabilities will allow mapping where the circumstellar matter is, and characterise how it reacts with and affects the star(s) and the planet(s) by imaging and characterising the physics with radiative transfer models. This will allow to constrain the two-way interplay of dusty host environments and planet formation and evolution.

% A high angular facility operating at the infrared wavelength regimes would allow for the imaging the the the warm dust emitted from circumstellar discs (planet forming discs around young stars and post-AGB discs) and stellar surfaces beyond what is reachable now. 

% While PLATO is expected to have found hundreds of Earth-like planets by the 2030s, the characterisation of the formation sites requires angular resolutions that we currently do not have, even for the nearest star forming regions.
% 

% umaging of the key processe

% embedded ...
% imaging
% link stages of stellar evolution, specifically the AGB and post-AGB stages. Both stages involve efficient dust grain processing
% mass distribution of solids

% Exoplanets as the outcomes of planet formation provide a zoo of end-
% points,

% UP TO HALF A PAGE:
% "A short, less than half a page, description of what technology developments/data handling requirements that may be needed can be included"
\vspace{-0.3cm}
\subsection{References}
\vspace{-0.3cm}
\footnotesize{
Andrews et al. 2018, ApJL 869, L41;
Andrych et al. 2025, PASA, 2, e125;
Bae et al. 2023, PPVII, ASPC 534, 423;
Bourdarot et al. 2020, A\&A, 639, A53; 
Bujarrabal et al. 2013,\,A\&A, 557, A104;
Close et al 2025 ApJL 990 L9;
Corporaal et al.\,2023,\,A\&A, 674, A151;
Coutu et al. 2019, ApJ, 885, 74;
Decin et al. 2020, Science, 369, 6510;
Defr\`ere et al. 2024, SPIE, 13095, 130950F;
De Ruyter et al. 2006,\,A\&A, 448, 641;
Drazkowska \& Alibert 2017, A\&A, 608, A92;
Flock et al. 2019; A\&A, 630, A147;
Gallardo Cava et al. 2021,\,A\&A, 648, A93;
GRAVITY+ Collaboration et al. 2025, A\&A, in press.
Haffert et al. 2019, Nat Astr. 3, 749;
Kennedy \& Kenyon 2008, ApJ, 673, 502;
Kluska et al. 2019,\,A\&A, 31, A108;
Kluska et al. 2022,\,A\&A, 658, A36;
Keppler et al. 2018, 617, A44;
Lau et al. 2024; A\&A, 688, A22;
Maes et al 2021, A\&A, 653, A25;
Maoi et al. 2025, A\&A, 699, L10;
Nielsen et al. 2023, A\&A, 678, A74;
Owen 2020, MNRAS, 495, 3160–3174,
Prisinzano et al. 2022, A\&A, 664, A175;
Rauer et al. 2014, Experimental Astronomy, 38, 249;
Rauer et al. 2025, Experimental Astronomy, 59, 26;
Soszy\'nski et al. 2021, ApJ, 911, L22;
van Capelleveen et al. 2025, ApJL 990 L8;
Veras et al. 2016, MNRAS, 458, 3942-3967;
Veras 2016, RSOS 3, 150571;
Wolff et al. 2016, ApJL, 818, L15;
Zuckerman et al. 2003, ApJ, 596, 477.
}

\end{multicols}

\end{document}